\documentclass[11pt,fleqn]{article}

\usepackage{amsmath,amssymb,amsthm,enumerate}

\newcommand{\sign}{\mathop{\rm sign}\nolimits}

\newcommand{\CL}{\mathop{\rm CL}\nolimits}

\newcounter{tbn}

\newcounter{mcasenum}

\newtheorem*{proposition*}{Proposition}
{\theoremstyle{definition}

}

\begin{document}

\par\noindent {\LARGE\bf
Group Analysis of Nonlinear Fin Equations
\par}

{\vspace{4mm}\par\noindent {\it O.~O.~Vaneeva$^{\dag 1}$, A. G.
Johnpillai$^{\ddag 2}$, R.~O.~Popovych$^{\dag\natural 3}$ and C.
Sophocleous$^{\S 4}$ }
\par\vspace{2mm}\par} {\vspace{2mm}\par\noindent {\it
${}^\dag$Institute of Mathematics of NAS of Ukraine, 3 Tereshchenkivska Str., Kyiv-4, 01601 Ukraine\\
$^\ddag$Department of Mathematics, Eastern University, Chenkalady, Sri Lanka  \\
$^\natural$Fakult\"at f\"ur Mathematik, Universit\"at Wien, Nordbergstra{\ss}e 15, A-1090 Wien, Austria\\
$^\S$Department of Mathematics and Statistics, University of Cyprus, Nicosia CY 1678, Cyprus\\
}}
{\noindent {\it $^1$vaneeva@imath.kiev.ua,
$^2$andrewgratienj@yahoo.com, $^3$rop@imath.kiev.ua,
$^4$christod@ucy.ac.cy\!}\par}

{\vspace{5mm}\par\noindent\hspace*{8mm}\parbox{146mm}{\small Group
classification of a class of nonlinear fin equations is carried
out exhaustively. Additional equivalence transformations and
conditional equivalence groups are also found. These allow to
simplify results of classification and further applications of
them. The derived Lie symmetries are used to construct exact
solutions of \emph{truly} nonlinear equations for the class under
consideration. Nonclassical symmetries of the fin equations are
discussed. Adduced results amend and essentially generalize recent
works on the subject [M. Pakdemirli and A.Z. Sahin, {\it Appl.
Math. Lett.}, 2006,  V.19, 378--384; A.H. Bokhari, A.H. Kara and
F.D. Zaman, {\it Appl. Math. Lett.}, 2006, V.19, 1356--1340].
}\par\vspace{2mm}}

\section{Introduction}

Investigation of heat conductivity and diffusion processes leads to interesting mathematical models
which can be often formulated in terms of partial differential equations.
In general case coefficients of model equations explicitly include both dependent and independent model variables
that makes difficulties in studying such models.

In this letter the class of nonlinear fin equations of the general form
\begin{equation}\label{EqFin}
u_t=(D(u)u_x)_x+h(x)u,
\end{equation}
where $D_u\not=0$, is investigated within the symmetry framework.
Here $u$ is treated as the dimensionless temperature,
$t$ and $x$ the dimensionless time and space variables,
$D$ the thermal conductivity, $h=-N^2f(x)$, $N$ the fin parameter and $f$ the heat transfer coefficient.

The condition $D_u=0$ corresponds to the linear case
of~\eqref{EqFin} which was completely investigated with the Lie
symmetry point of view long time
ago~\cite{Lie1881,Ovsiannikov1982}. Moreover, the sets of the
linear and nonlinear equations of form~\eqref{EqFin} can be
separately investigated under restriction with point symmetries.
That is why the linear case is excluded from consideration in the
present letter.

The problem of group classification for the degenerate case $h=0$
(i.e.\ the class of nonlinear one-dimensional diffusion equations)
was first solved by
Ovsiannikov~\cite{Ovsiannikov1959,Ovsiannikov1982}. The equations
of form~\eqref{EqFin} with $h$ being a constant are in the class
of diffusion--reaction equations classified by
Dorodnitsyn~\cite{Dorodnitsyn1982,Ibragimov1994V1}. Group
classification of the subclass where the thermal conductivity is a
power function of the temperature was carried out
in~\cite{Vaneeva&Johnpillai&Popovych&Sophocleous2006}. We keep the
above cases in presentation of results for reasons of
classification usability. Note also that Lie symmetries of the
class of quasi-linear parabolic equations in two independent
variables, which has a wide equivalence group and covers all the
mentioned classes, were classified
in~\cite{Basarab-Horwath&Lahno&Zhdanov2001,Lahno&Spichak&Stognii2002}.

Recently Lie symmetries and reductions of equations from
class~\eqref{EqFin} were considered in a number of
papers~\cite{Bokhari&Kara&Zaman2006,Pakdemirli&Sahin2004,Pakdemirli&Sahin2006}.
(See ibid for references on physical meaning and applications of
equations~\eqref{EqFin}.) In contrast to these papers, study in
our letter is concentrated on rigorous and exhaustive group
classification of the whole class~\eqref{EqFin} and construction
of exact solutions for \emph{truly} nonlinear
'variable-coefficient' equations from this class. Additional
equivalence transformations and conditional equivalence groups are
also found. These allow to simplify results of classification and
further applications of them. To find exact solutions, we apply
both classical Lie reduction and nonclassical symmetry approaches.

\section{Group classification and related problems}

Group classification of class~\eqref{EqFin} is performed in the
framework of the classical approach~\cite{Ovsiannikov1982}. All
necessary objects (the equivalence group, the kernel and all
inequivalent extensions of maximal Lie invariance algebras) are
found. Moreover, we extend the classical approach with additional
equivalence transformations and conditional equivalence group for
simplification of the classification results.

The \emph{equivalence group}~$G^\sim$ of class~\eqref{EqFin} is formed by the transformations
\[
\tilde t=\delta_1t+\delta_2,\quad \tilde x=\delta_3x+\delta_4,\quad \tilde u=\delta_5u,\quad
\tilde D=\delta_1^{-1}\delta_3^2D,\quad \tilde h=\delta_1^{-1}h,
\]
where $\delta_i$, $i=1,\ldots,5$, are arbitrary constants, $\delta_1\delta_3\delta_5\not=0$.
The connected component of the unity in $G^\sim$ is formed by continuous transformations having
$\delta_1>0$, $\delta_3>0$ and $\delta_5>0$.
The complement discrete component of~$G^\sim$ is generated by three
involutive transformations of alternating sign in the sets
$\{t,D,h\}$, $\{x\}$ and $\{u\}$.

The \emph{kernel} of the maximal Lie invariance algebras of equations from class~\eqref{EqFin}
coincides with the one-dimensional algebra $\langle\partial_t\rangle$.

All possible $G^\sim$-inequivalent \emph{cases of extension} of the maximal Lie invariance algebras are exhausted by ones
adduced in Table~1, where
\[
h^1(x)=\varepsilon\exp\left [\int \frac q{x^2+p}dx\right ];\quad
p\in\{-1,0,1\},\, \varepsilon=\pm1,\,\alpha\in\{0,1\}\!\!\!\mod G^\sim;\quad
n,\,q\neq 0.
\]

\begin{center}
\renewcommand{\arraystretch}{1.4}
\textbf{Table 1.} Results of group classification
\\[2ex]
\begin{tabular}{|c|c|c|l|}
\hline
N&$D(u)$&$h(x)$&\hfil Basis of $A^{\max}$ \\
\hline
1&$\forall$&$\forall$&$\partial_t$\\
\hline
2&$\forall$&1&$\partial_t,\  \partial_x$\\
\hline
3&$\forall$&$x^{-2}$&$\partial_t,\  2t\partial_t+x\partial_x$\\
\hline
4&$u^n$&$\varepsilon x^q$&$\partial_t,\  -qnt\partial_t+nx\partial_x+(q+2)u\partial_u$\\
\hline
5&$u^n$&$\varepsilon e^x$&$\partial_t,\  -nt\partial_t+n\partial_x+u\partial_u$\\
\hline
6&$u^{-4/3}$&$h^1(x)$&$\partial_t,\  -4qt\partial_t+4(x^2+p)\partial_x-3(4x+q)u\partial_u$\\
\hline
7&$\forall$&$0$&$\partial_t,\  \partial_x,\  2t\partial_t+x\partial_x$\\
\hline
8&$(u+1)^{-1}$&$\varepsilon$&$\partial_t,\  \partial_x,\
e^{\varepsilon t}\partial_t+\varepsilon e^{\varepsilon t}(u+1)\partial_u$\\
\hline
9&$e^u$&$0$&$\partial_t,\  \partial_x,\  2t\partial_t+x\partial_x,\  x\partial_x+2\partial_u$\\
\hline
10&$u^n,\  n\neq-\frac 43$&$\varepsilon$&$\partial_t,\  \partial_x,\
e^{-\varepsilon nt}(\partial_t+\varepsilon u\partial_u),\  nx\partial x+2u\partial_u$\\
\hline
11&$(u\!+\!\alpha)^n,\  n\neq-\frac 43$&$0$&
$\partial_t,\  \partial_x,\  2t\partial_t+x\partial_x,\  nx\partial_x+2(u\!+\!\alpha)\partial_u$\\
\hline
12&$u^{-4/3}$&$\varepsilon$&$\partial_t,\  \partial_x,\
e^{\frac 43\varepsilon t}(\partial_t+\varepsilon u\partial_u),\
2x\partial_x-3u\partial_u,\
x^2\partial_x-3xu\partial_u$\\
\hline
13&$(u\!+\!\alpha)^{-4/3}$&$0$&$\partial_t,\  \partial_x,\  2t\partial_t+x\partial_x,\
2x\partial_x-3(u\!+\!\alpha)\partial_u,\ x^2\partial_x-3x(u\!+\!\alpha)\partial_u$\\
\hline
\end{tabular}
\end{center}

The Case~6 was missed in~\cite{Pakdemirli&Sahin2004} and in subsequent papers on the subject.
The parameter-function $h^1$ equals to the following functions depending on values of~$p$:
\[
p=-1\colon\ h^1=\varepsilon\left|\frac{x-1}{x+1}\right|^{q/2}, \quad
p=0\colon\ h^1=\varepsilon e^{-q/x}, \quad
p=1\colon\ h^1=\varepsilon e^{q\arctan x}.
\]
Additionally we can assume $q=-1\!\!\!\mod G^\sim$ if $p=0$.

Some cases from Table~1 are equivalent with respect to point transformations which obviously do not belong to~$G^\sim$.
These transformations are called \emph{additional equivalence transformations}
and lead to simplification of further application of group classification results.
The pairs of point-equivalent extension cases and the corresponding additional equivalence transformations are
\begin{gather*}
6_{p=0}\to 5_{\tilde n=-4/3}\colon\quad \tilde t=t,\ \tilde x=x^{-1},\ \tilde u=x^3u;
\\
6_{p=-1}\to 4_{\tilde n=-4/3,\,\tilde q=q/2}\colon\quad
\tilde t=t,\ \tilde x=\frac{x-1}{x+1},\ \tilde u=2^{-3/2}(x+1)^3u;
\\[.5ex]
11_{\alpha\ne0}\to11_{\alpha=0},\ 13_{\alpha\ne0}\to13_{\alpha=0}\ (n=-\tfrac43)\colon\quad
\tilde t=t,\ \tilde x=x,\ \tilde u=u+\alpha;
\\[1.5ex]
10\to11_{\alpha=0},\ 12\to13_{\alpha=0}\ (n=-\tfrac43)\colon\quad
\tilde t=\tfrac1{\varepsilon n}e^{\varepsilon nt},\ \tilde x=x,\ \tilde u=e^{-\varepsilon t}u.
\end{gather*}
The latter transformation was adduced e.g.\ in~\cite{Ibragimov1994V1}.
Case~6 with $p=1$ is reduced to Case~4 only over the complex field.
Note also that Case~8 is reduced by the similar transformation
\[
\tilde t=-\tfrac1{\varepsilon}e^{-\varepsilon t},\ \tilde x=x,\ \tilde u=e^{-\varepsilon t}(u+1)
\]
to the equation~$\tilde u_{\tilde t}=(\tilde u^{-1}\tilde u_{\tilde x})_{\tilde x}-\varepsilon$
which is not in the class under consideration.

There are no other additional equivalence transformations in class~\eqref{EqFin}.
Therefore, up to point equivalence, possible cases of extension of maximal Lie invariance algebras are exhausted by
Cases~1--5, 6$_{p=1}$, 7--9, 11$_{\alpha=0}$ and 13$_{\alpha=0}$.

The singularity of the diffusion coefficient $D=u^{-4/3}$ with a number of different values of $h$ admitting
extensions of Lie invariance algebra can be explained in the framework of \emph{conditional equivalence groups}.
The equivalence group is extended under the condition $D=u^{-4/3}$.
More precisely, the equivalence group $G^\sim_1$ of the subclass of equations~\eqref{EqFin} with $D=u^{-4/3}$
is formed by the transformations
\[
\begin{array}{l}
\tilde t=\delta_1 t+\delta_2,\quad
\tilde x=\dfrac{\delta_3x+\delta_4}{\delta_5 x+\delta_6}, \quad
\tilde u=\pm\delta_1 (\delta_5 x+\delta_6)^3u, \quad
\tilde h={\delta_1}^{-1} h, \quad
\end{array}
\]
where $\delta_i$, $i=1,\ldots,6$, are arbitrary constants,
$\delta_1>0$ and $\delta_3\delta_6-\delta_4\delta_5=\pm1$.
$G^\sim_1$ is a non-trivial conditional equivalence group of class~\eqref{EqFin}.
Two first additional equivalence transformations belong to $G^\sim_1$.

Another example of a conditional equivalence group in class~\eqref{EqFin} arises under the condition $h=0$.
The equivalence group~$G^\sim$ of the whole class is then extended with translations with respect to~$u$,
i.e.\ the complete equivalence group $G^\sim_2$ of nonlinear diffusion equations ($h=0$)
is formed by the transformations
\[
\tilde t=\delta_1t+\delta_2,\quad \tilde x=\delta_3x+\delta_4,\quad \tilde u=\delta_5u+\delta_6,\quad
\tilde D=\delta_1^{-1}\delta_3^2D,
\]
where $\delta_i$, $i=1,\ldots,6$, are arbitrary constants, $\delta_1\delta_3\delta_5\not=0$.
The third additional equivalence transformation belongs to $G^\sim_2$.

The subclass of equations~\eqref{EqFin} with $h$ being a constant
admits an extension of generalized equivalence group. The prefix
``generalized'' means that transformations of the variables $t$,
$x$ and $u$ can depend on arbitrary elements~\cite{Meleshko1994}.
The associated generalized equivalence group~$G^\sim_3$ is
generated by transformations from~$G^\sim$ and the latter kind of
additional equivalence transformations, where $\varepsilon$ is
replaced by~$h$.

Knowledge on conditional equivalence groups allows us to describe the set of admissible (from-preserving)
transformations in class~\eqref{EqFin} completely.
See e.g.\ \cite{Vaneeva&Johnpillai&Popovych&Sophocleous2006} and references therein.

Note also that the subclass of equations~(\ref{EqFin}) possessing
nontrivial local conservation laws is exhausted by ones with $h$
is a constant. Then the corresponding space $\CL$ of conservation
laws is two-dimensional. The conserved vectors and the
characteristics of basis elements of $\CL$ are
\[\bigl(\,xe^{-ht}u,\ e^{-ht}\bigl(- xDu_x+{\textstyle\!\int\!\!D}\bigr)\,\bigr),\ xe^{-ht}
\quad\mbox{and}\quad
(\,e^{-ht}u,\ - e^{-ht}Du_x\,),\ e^{-ht}.
\]

\section{Similarity solutions}

Cases~7--13 of Table 1 are presented by `constant coefficient' diffusion--reaction equations.
Moreover, all of these cases either are usual nonlinear diffusion equations or can be reduced to them
by additional equivalence transformations.
Exact solutions of `constant coefficient' diffusion--reaction equations have been already investigated
intensively. See for example,~\cite{Dorodnitsyn1982,Galaktionov1990,Ibragimov1994V1,Polyanin&Zaitsev2004}.
That is why we choose Cases~4--6 as representatives among truly nonlinear variable-coefficient fin equations,
which are most interesting for Lie reduction.

As shown in the previous section, the equation
\begin{gather}\label{EqCase6}
u_t=(u^{-4/3}u_x)_x+h^1(x)u
\end{gather}
(Case~6 of Table 1)
admits the two-dimensional (non-commutative) Lie invariance algebra~$\mathfrak g$ generated by the operators
\[
X_1=\partial_t,\quad X_2=-4qt\partial_t+4(x^2+p)\partial_x-3(4x+q)u\partial_u.
\]
A complete list of inequivalent non-zero subalgebras of~$\mathfrak g$ is exhausted by the algebras
$\langle X_1\rangle$, $\langle X_2\rangle$ and $\langle X_1, X_2\rangle$.

Lie reduction of equation~\eqref{EqCase6} to an algebraic equation
can be made with the two-dimensional subalgebra $\langle X_1, X_2\rangle$
which coincides with the whole algebra~$\mathfrak g$.
The associated ansatz and the reduced algebraic equation have the form

\medskip

6.0. $\langle X_1,X_2\rangle$:\quad  $u=C(x^2+p)^{-3/2}\bigl(h^1(x)\bigr)^{-3/4}$,\quad
$C^{4/3}=\frac {3}{16}(q^2+16p)$.

\medskip

\noindent
Substituting the solution $C=\pm \frac {3^{3/4}}{8}(q^2+16p)^{3/4}$ of the reduced algebraic equation into the ansatz,
we construct the exact solution
\[
u=\pm \frac {3^{3/4}}{8}(q^2+16p)^{3/4}(x^2+p)^{-3/2}\bigl(h^1(x)\bigr)^{-3/4}
\]
of equation~\eqref{EqCase6}.

The ansatzes and reduced equations corresponding to the one-dimensional subalgebras from the optimal system
are following:

\bigskip

6.1. $\langle X_1\rangle$:\quad $u=(\varphi(\omega))^{-3}$, $\omega=x$;\quad
$3\varphi_{\omega\omega}=h^1(\omega)\varphi^{-3}$;

\bigskip

6.2. $\langle X_2\rangle$:\quad
$u=\bigl((x^2+p)^{1/2}(h^1(x))^{1/4}\varphi(\omega)\bigr)^{-3}$,
$\omega=th^1(x)$;

\medskip

$\qquad\qquad
3q^2{\omega}^2\varphi_{\omega\omega}+\frac 92q^2\omega\varphi_\omega-3\varphi^{-4}\varphi_\omega
+\frac 3{16}(q^2+16p)\varphi-\varepsilon\varphi^{-3}=0$.

\bigskip

The obtained reduced equations obviously have partial exact solutions which lead to
the above exact solution of equation~\eqref{EqCase6} and can be constructed via reduction to
algebraic equations. The problem is to find some different solutions.
We are only able to reduce the order of equation 6.1. Namely, in the variables
\[
y=(\omega^2+p)^{-1/2}(h^1(\omega))^{-1/4}\varphi,\quad
\psi=(\omega^2+p)^{-1/2}(h^1(\omega))^{-1/4}((\omega^2+p)\varphi_\omega-\omega\varphi)
\]
constructed with the induced symmetry operator $4(\omega^2+p)\partial_\omega+(4\omega+q)\varphi\partial_\varphi$
equation 6.1 takes the form
$
(4\psi-qy)\psi_y+q\psi+4py=\frac43\varepsilon y^{-3}.
$
A better way for construction of exact solutions for equations of Case~6 with $p\leqslant0$
is to map them to Cases~4 and~5 with additional equivalence transformations and then study the latter cases.

Let us review results on Lie reduction of Cases~4 and~6.
For each from these cases we denote the basis symmetry operators adduced in Table~1 by $X_1$ and~$X_2$.
Structure and list of inequivalent subalgebras of the Lie invariance algebras are the same as ones in Case~6.
The associated ansatzes and reduced equations have the form ($\varepsilon'=\sign t$):

\medskip

4.0. $\langle X_1,X_2\rangle$:\quad $u=Cx^{\frac{q+2}n}$,\quad
$(q+2)(nq+n+q+2)C^{n+1}+\varepsilon n^2C=0$;

\medskip

4.1. $\langle X_1\rangle$:\quad
$u=(\varphi(\omega))^{\frac1{n+1}}$,\ $\omega=x$,\quad
$\varphi_{\omega\omega}+\varepsilon(n+1)\omega^q\varphi^{\frac1{n+1}}=0$\quad if $n\ne -1$;\\[1ex]
$\phantom{\text{4.1. $\langle X_1\rangle$:}}$\quad
$u=\exp(\varphi(\omega))$,\ $\omega=x$,\quad
$\varphi_{\omega\omega}+\varepsilon\omega^qe^\varphi=0$\quad if $n=-1$;

\medskip

4.2. $\langle X_2\rangle$:\quad
$u=|t|^{-\frac{q+2}{nq}}\varphi(\omega)$,\ $\omega=|t|^{\frac1q}x$,\quad
$(\varphi^n\varphi_\omega)_\omega+\varepsilon\omega^q\varphi
+\varepsilon'\dfrac{q+2}{nq}\varphi-\varepsilon'\dfrac1q\omega\varphi_\omega=0$;

\bigskip

5.0. $\langle X_1,X_2\rangle$:\quad $u=Ce^{\frac xn}$,\quad
$(n+1)C^{n+1}+\varepsilon n^2C=0$;

\medskip

5.1. $\langle X_1\rangle$:\quad
$u=(\varphi(\omega))^{\frac1{n+1}}$,\ $\omega=x$,\quad
$\varphi_{\omega\omega}+\varepsilon(n+1)e^\omega\varphi^{\frac1{n+1}}=0$\quad if $n\ne -1$;\\[1ex]
$\phantom{\text{4.1. $\langle X_1\rangle$:}}$\quad
$u=\exp(\varphi(\omega))$,\ $\omega=x$,\quad
$\varphi_{\omega\omega}+\varepsilon e^{\varphi+\omega}=0$\quad if $n=-1$;

\medskip

5.2. $\langle X_2\rangle$:\quad
$u=|t|^{-\frac1n}\varphi(\omega)$,\ $\omega=x+\ln|t|$,\quad
$(\varphi^n\varphi_\omega)_\omega+\varepsilon e^\omega\varphi
+\varepsilon'n^{-1}\varphi-\varepsilon'\varphi_\omega=0$.

\bigskip

Reduction to algebraic equations gives the following solutions of the initial equations:

\medskip

4. $u=\left(-\dfrac{q+2}{\varepsilon n^2}(nq+n+q+2)\right)^{-\frac1n}x^{\frac{q+2}n}$;

\medskip

5. $u=\left(-\dfrac{n+1}{\varepsilon n^2}\right)^{-\frac1n}e^{\frac xn}$.

\medskip

There are Emden--Fowler and Lane--Emden equations and their different modifications
among the reduced ordinary differential equations. Solutions of these equations are known
for a number of parameter values (see e.g.~\cite{Polyanin&Zaitsev2003}).
As a result, classes of exact solutions can be constructed for fin equations corresponding to
Cases~4 and~5 of Table~1 for a wide set of the parameters $n$ and~$q$.

\section{On nonclassical symmetries}

We also study conditional (nonclassical) symmetries of equations from class~\eqref{EqFin}.
As well-known, the operators with the vanishing coefficient of $\partial_t$ gives so-called `no-go' case
in study of conditional symmetries of an arbitrary $(1+1)$-dimensional evolution equation
since the problem on their finding is reduced to a single equation which is equivalent to the initial one
(see e.g.~\cite{Zhdanov&Lahno1998}).
Since the determining equation has more independent variables and, therefore, more freedom degrees,
it is more convenient often to guess a simple solution or a simple ansatz
for the determining equation, which can give a parametric set of complicated solutions of the initial equation.
For example, the fin equation
\begin{equation}\label{EqFin-1x}
u_t=(u^{-1}u_x)_x+xu
\end{equation}
is conditionally invariant with respect to the operator $\partial_x+tu\partial_u$.
The associated ansatz $u=e^{tx}\varphi(\omega)$, $\omega=t$, reduces equation~\eqref{EqFin-1x}
to the equation $\varphi_\omega=0$,
i.e.\ $u=Ce^{tx}$ is its non-Lie exact solution which can be additionally extended with symmetry transformations.

It is known that
non-linear diffusion equations (i.e.\ equations~\eqref{EqFin} with $h=0$) possess conditional symmetry operators
which have non-vanishing coefficients of~$\partial_t$ and are inequivalent to Lie invariance operators
only in case of exponential diffusion coefficients
(Case~9 of Table~1 and equivalent equations). Moreover,
solutions associated with such conditional symmetry operators are still Lie invariant.
The opposite situation is for $h\not=0$.
There exist conditional symmetry operators of equations~\eqref{EqFin},
which have non-vanishing coefficients of~$\partial_t$, are inequivalent to Lie invariance operators and
even lead to truly non-Lie exact solutions.
Thus, consider again equation~\eqref{EqFin-1x}.
It admits also the conditional symmetry operator $\partial_t+xu\partial_u$.
The associated ansatz $u=e^{tx}\varphi(\omega)$, $\omega=x$, reduces equation~\eqref{EqFin-1x}
to the equation $(\varphi^{-1}\varphi_\omega)_\omega=0$.
The general solution $\varphi=C_1e^{C_2x}$ of the reduced equations gives a solution of~\eqref{EqFin-1x},
which is simplified to the above constructed one with symmetry transformations.

Exhaustive description of nonclassical symmetry operators of equations~\eqref{EqFin}
will be a subject of a forthcoming paper.

\section*{Acknowledgements}
The research of R.\,P. was supported by Austrian Science Fund
(FWF), Lise Meitner project M923-N13. The research of O.\,V. was
partially supported by the grant of the President of Ukraine for
young scientists GF/F11/0061. A.\,G.\,J. would like to thank Eastern
University, Sri Lanka for its continuing support in his research
work. R.\,P., O.\,V. and A.\,G.\,J. are grateful for the hospitality
and financial support by the University of Cyprus.

\end{document}